\documentclass[twocolumn,aps,prd,amsmath,amssymb]{revtex4}
\bibpunct{}{}{,}{s}{}{,}
\usepackage{graphicx}
\newcommand{\be}{\begin{equation}}
\newcommand{\ee}{\end{equation}}
\newcommand{\bea}{\begin{eqnarray}}
\newcommand{\eea}{\end{eqnarray}}
\begin{document}
\title{Comments on Background Independence and Gauge Redundancies}
\author{Moshe Rozali}
\email{rozali@phas.ubc.ca}
\affiliation{University of British Columbia}

\begin{abstract}
  We describe the definition and the role background independence and the closely related notion of diffeomorphism invariance play in modern string theory. These important concepts are transformed by a new understanding of gauge redundancies and their implementation in non-perturbative quantum field theory and quantum gravity. This new understanding also suggests a new role for the so-called background-independent approaches to directly quantize the gravitational field. This article is intended for a general audience, and is based on a plenary talk given in the Loops 2007 conference in Morelia, Mexico.

\end{abstract}

\maketitle

\section{Introduction}

The problem of quantum gravity is one of the biggest remaining mysteries in physics. Many conceptual and technical issues make it difficult to treat the gravitational field quantum mechanically in the same way one quantizes, for example, the electromagnetic fields. Moreover, one is not likely to get guidance from experimental results, since the energy scale associated with quantum gravitational effects is enormous, around $10^{19} GeV$. In those circumstances one has to turn to theoretical considerations and consistency checks, to narrow down the range of possibilities.

One of the ideas that has provided  valuable clues to the nature of quantum gravity is that of {\it Background Independence}. Recently this rather technical concept has captured wide interest well beyond the quantum gravity community, spreading into the
general physics community and beyond. Furthermore, the ideas of duality and holography, developed in the last decade, have altered  the role these concepts play in quantum gravity research. The purpose of this article, and the talk it is based on, is to provide a personal perspective on this rather nebulous notion, and the closely related notion of diffeomorphism invariance (and more generally gauge redundancies), from the perspective of a string theorist.
The article is based on knowledge common to most practitioners of modern quantum field theory and string theory,  which could be found in many review articles and textbooks, therefore the list of references will be far from exhaustive.
Some references touching on similar issues are \cite{Banks:2000zy, seiberg, polchinski}.

The outline of this paper is as follows: this introductory section is devoted to defining the notion of background independence (henceforth BI) and motivating that definition. We  then turn to discussing the evolution of this concept in a more or less historical fashion, starting from perturbative string theory (which is argued to be background-dependent), going through an interlude regarding gauge invariance and dualities in quantum field theories, and ending with a fully BI example of non-perturbative string theory, in the form of the AdS/CFT duality \cite{Maldacena:1997re}. We end with a short discussion of future directions.

\subsection {Basic Definitions}

Consider the simplest classical or quantum mechanical system, that of a one-dimensional particle moving in the potential
$V(x)$. The system is described by a configuration space (parametrized by $x$) and our role as physicists is to describe the state of the system (classically the location $x$, or quantum mechanically the wavefunction $\psi(x)$) and its time evolution.

Oftentimes we have only an incomplete knowledge of $V(x)$, or we are only able to calculate the dynamics in the vicinity of some location $x_0$. In that case we call $x_0$ the "background", and we proceed by working in perturbation theory around that background. Obviously, this gives us only partial information. For example we are unable to find what is the ground state wavefunction of the system. More technically, the perturbation series is generally not summable, therefore it is not sufficient to extract information about large fluctuations away from the chosen background, or long time evolution of the system.

In quantum mechanics the separation to background and small fluctuations is related to the process of "quantization", the systematic inclusion of small quantum corrections around mostly classical background \footnote{An unfortunate but common usage of the word "quantize" is as a synonym for "making discrete". Discrete spectra played an important historical role in the development of quantum theory, but ultimately there is nothing fundamentally discrete about quantum mechanics.}. The background can be viewed roughly as a condensate (or a coherent state) of a large number of quanta, and including fluctuations around it account for the effect of additional finite number of quanta.

In the  case we separate the configuration of the system into classical and quantum parts, we call the description of the system "background dependent" and agree it is an incomplete state of knowledge. As described below, this is the situation for example for string perturbation theory. As described, the flaw is generally related to the use of approximate, perturbative techniques.

This definition can be generalized to more complicated systems as well. Supposed we are describing the electric and magnetic fields in flat space. Then Maxwell equations are a complete BI description of the system: the configuration space here is the infinite dimensional space of all possible field configurations $\vec{E}(x), \vec{B}(x)$. Maxwell's equations do not single out any  particular such configuration  and therefore it is BI, at least with the definition given here.

 With this definition of background, and BI, the problem of background dependence is not merely a philosophical or aesthetic unease. Rather, any question which requires comparison of different (generally vastly different) backgrounds, or any physical quantity that receives contribution (at the required precision level) from well-separated points in configuration space, any such physical issue will be inadequately addressed using a background-dependent formulation.
Examples of such issues, for which BI formulation is needed are:

 \begin{itemize}

 \item In quantum mechanics of a single particle in a double well-potential, perturbation theory around any of the minima of the potential will miss tunneling effects. Therefore, the ground state wave function cannot be correctly described using perturbation theory around a fixed background.

 \item In quantum electrodynamics, a constant electric field can decay by generating electron-positron pairs from the vacuum, which then screen the electric field. This is the famous Schwinger effect. Any description which singles out a particular configuration of the electric field will be inadequate in describing this effect.

 \item Moving to quantum gravity: with one's favorite cosmological selection principle, if such a principle exists, one could try to explain features of the observable universe or its initial conditions. Clearly this task requires comparison of diverse set of possible cosmologies, and dynamical transitions between them, and therefore a BI formulation.

\item Resolution of the information paradox seems to require contribution from separate backgrounds \cite{eternal,hawking}, and therefore a BI formulation of the problem.

 \end{itemize}

So, BI in the sense defined above seems like a desirable, indeed a necessary ingredient of any future theory of quantum gravity. One therefore often hears the sentiment that any such theory "must" be BI. Since I argue below that some definitions of string theory are already BI in the sense defined here, I now turn to motivating the above definitions and exemplify them using the paradigm of classical general relativity.

\subsection{Backgrounds versus Superselection Sectors}

The above definition of BI is in the spirit of the background field method of quantizing field theories (including gravitational theories). One fixes a background for all the fields (say a background metric if quantizing general relativity), and quantizes the small fluctuations around that background.  Normally, one would not call the results of this  method background dependent: all physical results do not depend on the background chosen, though the intermediate steps to obtaining those results might. Nevertheless, when working in perturbation theory, one can only show that BI holds when changing the background by a small (infinitesimal) amount. In other words the results are BI only in as much as they are perturbative, the question of full BI can only be addressed in the context of non-perturbative physics.

There are other notions of BI in the literature, but many of them reduce to the above upon closer inspection, whereas some other definitions are too narrow, essentially only applying to general relativity and attempts to directly quantizing it, but not to a more general approaches to quantum gravity. Yet some other definitions are so ambitious as to classify any existing or conceivable future theory as background-dependent. To exemplify precisely what is meant by the definition of BI given here, and what is not, I now turn to classical general relativity, which is often given as the paradigm of a BI theory.

Gravitational physics, including for example celestial mechanics, was described prior to Einstein's theory of general relativity by Newtonian mechanics.
In Newtonian mechanics the spatial and temporal coordinates of a star are an absolute concept, defined with respect to background spacetime. This raises some aesthetic and philosophical unease, described for example in \cite{Smolin:2005mq}, which is resolved by Einstein's promotion of the metric to a fluctuating field.
In Einstein's general theory of relativity the background now acquires a dynamical nature, it is determined by Einstein's equations and indeed can change with time. The description therefore is more economical, no background structure exists, it is determined dynamically.

In the context of the definitions above, the background discussed is the metric structure, which is fixed in Newtonian mechanics (or even in special relativity), but becomes dynamical in general relativity. This is a beautiful example of  replacing a background structure (which constitutes an arbitrary choice) by a dynamically chosen quantity (which then has a rationale and can be derived from more fundamental structure). This has led some to judge success in formulating a fundamental theory in the degree to which it is BI.  For example in the words of \cite{Smolin:2005mq}, the "relational strategy"\footnote{The degree to which general relativity, or indeed any local theory, can be "relational"  or "Machian" is not clear to me, see for example \cite{Barbour:1995iu} for interesting discussions. However, the issue is tangential to the topic of this paper. } is to "seek to make progress by identifying the background structure in our theories and removing it, replacing it with relations that evolve subject to dynamical law".

The usefulness of this strategy clearly depends on what constitutes a  "background" and what does not. Is there a useful and a-priori (theory-independent) definition of what constitutes a background and what does not?  in other words, how does one go about "identifying" a background?

  Following \cite{ss,Banks:2000zy} I suggest to define a background as a quantity that can change dynamically in an ordinary physical process (i.e. one that takes finite amount of time and involves finite amount of energy). A quantity that cannot be changed dynamically cannot be thought as a background, rather it is a fixed parameter of the theory, which defines what conventionally is called a superselection sector.

To drive the distinction home let me illustrate the difference in some familiar examples. The location of a single particle can change dynamically, therefore any formulation of the dynamics which singles out a specific location would be inconvenient or inadequate in describing processes in which the particle travels great distances.  We call such difficulty background-dependence. On the other hand, in non-relativistic quantum mechanics, the fact that we have a single particle (and not a few) is not changeable by a dynamical process. The process of identifying and removing the background in this context (conventionally known as second quantization) will make new physical processes possible, for example such processes in which the net number of particles changes with time.

In this way, Einstein's generalization of Newtonian gravity involves the realization that some aspects of the geometry of spacetime are dynamical, and therefore the fixed background of Newtonian dynamics is inadequate description for those physical processes- namely exactly those in which spacetime geometry changes (for example a collapse to a black hole  or even the more mundane passage of gravitational wave through a detector).

However, it is important to note that in all those examples there are some aspects of the theory that stay non-dynamical, and cannot sensibly be considered a "background". For example in general relativity, the form of the asymptotic geometry is one such aspect.
The phase space of classical general relativity in asymptotically flat space is different from the one of asymptotically anti-de-Sitter (AdS) space. More physically - there is no finite process in classical general relativity which converts asymptotically flat space to asymptotically AdS one.  Those two theories, which  are  distinguished by choice of asymptotic geometry (or more generally boundary conditions),  should be thought of as defining superselection sectors, they are really different, and are not related by changing the "background" in some more fundamental theory.

This situation is not at all unusual, in all conventional theories dynamics is described by  a set of differential equations, and the set of solutions depends on the choice of boundary conditions. Those boundary conditions do not change dynamically, and cannot be considered to be a background. Rather, they are part of the data needed to specify the dynamical problem. It is only with this distinction between background and superselection sectors, that classical general relativity, or indeed any other physical theory, can be considered to be BI. In order to have a meaningful discussion then, I will  adopt this definition of what is, and is not, a background. As we shall see below, existing holographic formulations of string theory in various circumstances are fully BI in that sense.

\subsection{BI and Gauge Redundancies}

Once we consider the quantum mechanics of the electric and magnetic fields, or that of the gravitational field, we have a new and interesting complication, that of gauge freedom \footnote{ As explained below, the term "gauge symmetry" is a misnomer to be avoided. Unlike symmetries, gauge redundancies have no physical consequences.}. Classically, one introduces potentials $\vec{A}, \Phi(x)$ in order to simplify the equations, but they are not necessary in principle, Maxwell equations are already a complete description of the system. However, as demonstrated for example by the Aharonov-Bohm effect, quantum mechanically the situation is different. There are aspects of the system (such as the interference pattern of electrons in the presence of a localized magnetic field) which are insensitive to local values of the electric and magnetic fields alone. Rather, they are summarized by global observables (holonomies, or Wilson loops). In terms of the original fields, those holonomies would be non-local observables. The introduction of potentials is necessary to restore manifest locality to our description of the system.

The price to pay is that of gauge freedom: with the introduction of potentials we have now many different potentials which encode the same physics. If we denote the space of all configurations $\vec{A}(x), \Phi(x)$ by $C_L$ (L stands for "large"), the physical configuration space of the system is much smaller. One has to account for the fact that many configurations in $C_L$ are physically identical and therefore the real configuration space is schematically $C_L/G$, where $G$ denotes symbolically the identifications we have to impose due to gauge invariance.

We see therefore that gauge invariance is distinct, but intimately related, to BI. Traditionally, gauge invariance is achieved by working in the larger configuration space and imposing constraints ensuring that physical quantities are gauge invariant. This goes a long way towards ensuring BI as well. One of the lessons of modern investigations of non-perturbative QFT and string theory is that gauge redundancies (including diffeomorphism invariance) are not fundamental and are tied  inherently to a particular perturbative expansion of the theory. As such they are inherently background-dependent. I will describe this below, and as we will see in several examples, this also has implications for the best strategy to achieve BI in a physical theory.

\section{Background Dependence in Perturbative String Theory}

One of the most common  mental images of string theory is that of the string worldsheet, the surface spanned by  a string as a function of time, in a fixed spacetime manifold. The string worldsheet is then a map $X:\Sigma \rightarrow M$ between a two-dimensional surface $\Sigma$ and a fixed spacetime manifold $M$, parametrized locally by coordinates $X$, and is endowed with the metric $G_{\mu\nu}(X)$. One describes the "quantization" of the string\footnote{First quantization of point particles corresponds to the transition from classical particles to classical fields. Similarly quantization of the string in this context defines classical string theory, quantum effects are included perturbatively by considering the worldline or worldsheet theory on higher genus graphs or surfaces.} by summing over all such worldsheets with prescribed boundary conditions, and worldsheet action of the form
\begin{equation}
I_{ws}= \frac {1}{4\pi \alpha'}\int d^2\sigma \,\partial^a X^\mu \partial_b X^\nu G_{\mu\nu}(X)
\end{equation}
where the worldsheet is parametrized by $\sigma^a, a=1,2$, and $\alpha'$ is the inverse string tension.  The manifold $M$ with the metric $G_{\mu\nu}(X)$  are a "background" of string theory, namely a manifold on which strings can be consistently quantized. One of the mysterious and exciting  results in perturbative string theory is the fact that the string can be consistently quantized if and only if the metric satisfies Einstein's equation (with calculable higher derivative corrections suppressed in low energies).

Despite being a very common mental image, and the historical starting point of the subject, this picture is misleading in some important ways. Some of them related to the issue of background independence. I will describe the situation briefly, since the main purpose here is to concentrate on non-perturbative physics.

First, in addition to background spacetime, one has to choose in general backgrounds for all the other massless modes of the string. The conditions for consistency of the string propagation (absence of negative norm states) then relate those backgrounds by a set of differential equations including the Einstein equation coupled to matter. In this sense the string background is a generalization of the background used in quantizing quantum field theories via the conventional background field method. The diffeomorphism symmetry is manifested precisely as it does in background field quantizations of gravity or gauge theories. One can show explicitly (and quite easily) that the formulation is diffeomorphism invariant with respect to infinitesimal diffeomorphisms, which is all one can expect in a perturbative framework.

It is also worth noting that the so-called sigma model action given above, describing a propagation in weakly curved spacetime with slowly varying fields, is {\it not} the most general string background. Rather, the general perturbative string background is described by a two dimensional conformal field theory. Most of those backgrounds do not resemble a classical spacetime at all, they are abstract string backgrounds with no geometrical interpretation. Some subset of possible string backgrounds resemble classical spacetime only in some limit, when a parameter is tuned to an extreme value (in those circumstances the parameter is interpreted as a size of a geometrical feature of spacetime, which becomes large in the limit).  Moreover, many times there is more than one such spacetime interpretation for a given string background. In this sense spacetime is inherently a derived concept in string theory, even perturbatively. Any relevant concept, including that of BI, has to avoid explicit reference to spacetime structures in order to be applicable in this context.

However, quantizing the string perturbatively is clearly not a complete description of the physics, and there are many examples of interesting questions which require a more complete description. Before jumping into non-perturbative string theory, I will make a brief detour into non-perturbative gauge theories, to discuss the important idea of duality.

\section{Interlude: Field Theory Dualities and Gauge Invariance}

Before returning to quantum gravity, the main subject of this article, let us demonstrate the role of gauge invariance in the simpler context of quantum field theory. We discuss the case of strongly coupled non-Abelian gauge theory, and concentrate on sufficiently low energies, where the theory flows to an interacting conformal field theory. This example is chosen for pedagogical reasons, as one of the simplest instances of duality, but most of its specific properties are not important. The phenomena of duality is generic in quantum field theories, and as we will see next the same set of ideas applies (in all energy scales) to more complicated examples involving quantum gravity.

In order to gather evidence for duality, one needs to make exact non-perturbative calculations, or to make qualitative arguments. The former is possible in a special set of theories, and the latter gives confidence the phenomena discovered are generic. For the purpose of illustration I'll concentrate on four dimensional theories with a single supersymmetry. The duality exhibited at low energies is known as Seiberg duality \cite{duality}.

So, let us consider an $SU(n_c)$ gauge theory with $n_f$ chiral multiplets in the fundamental representations ("flavors"). Let us call that formulation "description A" of the theory, shortly we will discuss another description which is equivalent. To be in the regime where the theory flows to a non-trivial CFT at low energies, we need to restrict the range of $n_f,n_c$ appropriately, let us do so.

When quantizing theory A in perturbation theory one constructs the Hilbert space from the fields in the action: quarks and gluons (and their supersymmetric partners). Let us denote the resulting space by $H_L$ (the large Hilbert space).  The space $H_L$ is not really a Hilbert space, it has negative norm states, thus the need for gauge invariance. Gauge invariance can be implemented in different ways (e.g BRST quantization), in all of them one restrict attention to a smaller Hilbert space, one on which the constraints of gauge invariance have been consistently imposed. Let us call the reduced Hilbert space $H_S$, the small Hilbert space.

However, even at weak coupling the spectrum is much richer, and the physical Hilbert space is bigger that just $H_S$, including for example solitonic excitations whose mass scales as inverse powers of the coupling constant (so they become infinitely heavy in the classical limit). It is not clear what  role, if any, the original Hilbert space $H_L$ and the the gauge constraint, have in the full  theory, away from the weak coupling region where perturbation theory applies. After all, the states in the physical Hilbert space are precisely those which are invariant under the constraints, and all physical quantities are gauge invariant. Gauge invariance, by construction, has no physical consequences.

Those semi-philosophical concerns become more urgent due to the discovery of duality symmetries. It turns out, in a growing number of examples, that one can quantize different gauge theories, which look very different in perturbation theory, yet obtaining precisely the same non-perturbative physics. Conversely, one can have non-perturbative quantum field theories which have more than one weak coupling limit. In each such limit they look like some weakly coupled gauge theory, but the details - the matter content, the gauge redundancies, the Lagrangian, are different in each limit.

So, in the case of Seiberg Duality discussed here, we have an equivalent description, theory B. That description  involves an $SU(n_f-n_c)$ theory, with $n_f$ flavors and one additional scalar field $M$ (which is a gauge singlet), and with specific interactions.  The new gauge theory looks very different from the original one, it utilizes different variables (fields) and has different gauge redundancies, however it turns out that all the non-perturbative physics (at low energies, for the range of $n_f,n_c$ discussed above) is exactly identical!

The situation is described by the following diagram

\begin{figure}[htp]
\centering
\includegraphics[width=85mm]{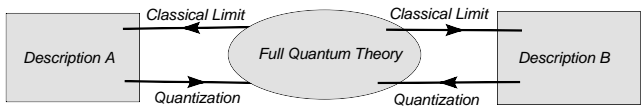}
\caption{Quantum Theory possessing multiple classical limits, which can be obtained by quantizing any one of them and increasing the coupling constant.}\label{fig:erptsqfit}
\end{figure}

The full theory has two limits in which it simplifies. When some coupling is taken to an extreme value (usually chosen to be called then "weak coupling"), the theory starts looking like theory A. Results near that limit, in which the coupling is weak, are reliably obtained by "quantizing" theory A and treating it perturbatively. Similarly, the same theory has another limit in which another coupling is taken to be weak (perhaps the inverse of the original coupling), where it reduces in the same sense to theory B.

The discovery of duality makes it necessary to distinguish between concepts that are well-defined and useful non-perturbatively, and concepts that are specific to a certain classical limit, and the set of variables best suited for that limit. It turns out that the set of fundamental fields, their Lagrangian and the associated gauge redundancies are all specific to a choice of variables, and are not intrinsic properties of the full non-perturbative theory.

Let us look more closely at the gauge redundancies of both descriptions, for the example of Seiberg duality given above. In the first description the gauge invariance $SU(n_c)$ is realized the traditional way, for example by quantizing canonically and imposing Gauss law constraints. On the other hand, in the set of variables utilized in description B, the original gauge freedom $SU(n_c)$ is invisible. In other words the original $SU(n_c)$ gauge symmetry has different implementation in those variables, namely they are all singlets. In that sense description B utilizes gauge invariant variables, albeit at the cost of introducing a new gauge redundancy (which, in turn, is invisible in the first set of variables). It is interesting that the gauge invariant variables tend to have their own gauge redundancy (though that is not always the case). I'll make some more comments on that phenomena in the conclusions.

\section{Non-Perturbative Gravity: Duality and Holography}

We have seen that quantum field theories, in various dimensions and with various amount of supersymmetry, have the property of duality. This means they can be described in many equivalent ways, or in other words using many different variables. Each description, or set of variables, with all the associated mental imagery, is closely tied to a particular classical limit of the theory. The gauge redundancy is a facet of the description, not an intrinsic property of the physics. Different descriptions have different redundancies, but the same physics, which is by definition independent of all those redundancies.

What about quantum gravity? can we exhibit similar behavior in gravitating systems? the answer is yes. There are now many examples of gravitational theories which have a  more conventional dual description, that of a lower-dimensional non-gravitational theories.  In other words, there exist examples of quantum theories which posses more than one classical limit, and in one or more of those limits they look like weakly interacting gravitational theories, whereas in other limits they look non-gravitational. In fact holographic dualities seem ubiquitous, any known non-perturbative formulation of quantum gravity seems to be related to lower dimensional field theory, which lives in some vague sense on the boundary of spacetime, encoding holographically all the information in the bulk.

The most familiar of those holographic dualities  is the  AdS-CFT correspondence \cite{Maldacena:1997re}, which
 establishes a precise dictionary between all observable quantities in asymptotically AdS spaces and all physical quantities in a specific quantum field theory (the $N=4$ supersymmetric gauge theory).
This section is devoted to exploring the AdS/CFT correspondence  and its implications for BI and the role of diffeomorphism invariance in quantizing gravity. We start by defining the AdS/CFT correspondence precisely, continue by  describing some of its salient features, and then elaborate on the role of gauge invariance and the meaning of BI in this context.

 \subsection{Basics of AdS/CFT}

 Consider your favorite model of quantum gravity in asymptotically AdS space\footnote{We fix the boundary conditions, but not the background, $AdS$ space itself should be considered to be just one state, the vacuum state, of the theory. More on that in the main text.}. Five dimensional Anti de Sitter space is given by the metric (in global coordinates) \be ds^2 = -\cosh^2 \rho \,dt^2 + d\rho^2 +  \sinh^2 \rho \,d\Omega_3^2  \ee where $t$ denotes the global time coordinate, $d\Omega_3^2$ is the metric on the 3-sphere, and $\rho$ is a radial coordinate spanning half-line. The spacetime is  distinguished by having a timelike boundary, at $\rho=\infty$, which is conformal to $R \times S^3$. Therefore in order to completely specify the model one has to specify appropriate boundary conditions for all propagating degrees of freedom.  One can then discuss the physical observables of the theory as a function of those boundary conditions.

 Let us specialize to string theory, defined on asymptotically AdS spaces. Denote the set of string fields schematically by $\Phi(\rho, t, \Omega_3)$, where $\Omega_3$ stands for the angular coordinates of the 3-sphere. Each such field satisfies boundary conditions\footnote{Note that those boundary conditions can break all global symmetries, be time dependent, etc.} at $\rho=\infty$ which are given by a function $J(\rho, \Omega_3)$ (precise details on specifying those boundary conditions can be found at \cite{Witten:1998qj}). The complete information on the theory is encoded in  all observable quantities as function of the boundary conditions $J(\rho, \Omega_3)$.

 It is strongly believed that the only diffeomorphism invariant quantities are global observables, given by integrals over local densities (no counter-example to this claim is known). They are all encoded in the partition function
\be Z(J) = \int Dg...e^{i(S_{EH}(g)+...)}\ee
where the integral sign denotes symbolically some appropriately defined path integral, or stringy generalization thereof, which is used in quantizing the gravitational theory with the specified boundary conditions for all string fields. The quantity $Z(J)$ encodes all the gauge invariant quantities in asymptotically AdS space, quantizing string theory (which includes gravity) in asymptotically AdS spaces amounts to calculating the object $Z(J)$.

Let me elaborate on this point. The object $Z(J)$ encodes all the well-defined quantities in asymptotically AdS spaces, therefore it encodes all the answers to the  interesting questions regarding the  combination of quantum mechanics and gravity in such spaces. For example one can form small black holes and let them evaporate, perhaps even sending some observer through the apparent horizon in the process. All the well defined questions regarding this process  are contained in $Z(J)$, but not always in a manner easy to decode. Moreover, for small enough cosmological constant, any local processes in asymptotically AdS space is indistinguishable from the same process in asymptotically flat space. Only global issues, important for example for cosmology, would be sensitive to the difference in asymptotic boundary conditions. The calculation and interpretation of $Z(J)$ is of clear importance for anyone interested in quantum gravity.

 We now turn to the dual description. It turns out that the object $Z(J)$ can be calculated with no reference to quantizing gravity or AdS space. Consider the gauge theory mentioned above ($N=4$ SYM). The complete information about the gauge theory is encoded in all correlation functions of gauge invariant operators. This information is summarized in the object $Z(J)$, the partition function with sources. Schematically  \be Z(J)= \int DA...e^{i(S_{YM} +...+\int J \Theta)}\ee  where the integral sign stands for path integral over the non-Abelian gauge fields (in an $SU(n)$ adjoint representation) and their supersymmetric partners, weighted by the Yang-Mills action $S_{YM}$ (and additional terms involving the fermions and scalars). The sources  $J(\rho, \Omega_3)$ couple to  all local gauge invariant operators, denoted schematically by $\Theta$  (for example $\Theta (\rho, t, \Omega_3) = Tr(F_{\mu\nu} F^{\mu\nu}(\rho, t, \Omega_3) $). The partition function $Z(J)$ is a generating functional for all correlation functions of those local operators, which are obtained from $Z(J)$ by repeated differentiation.

 Witten's definition \cite{Witten:1998qj} of the AdS/CFT correspondence is simply the statement that the two functionals $Z(J)$ defined above are in fact one and the same. Calculating $Z(J)$ using the gauge theory variables amounts then to a complete non-perturbative, background-independent quantization of gravity in asymptotically AdS spaces.

 \subsection{Background Independence and Diffeomorphism Invariance}

 In specifying the AdS/CFT correspondence, we restrict attention to asymptotically AdS spaces. On the gauge theory side of the correspondence, we restrict to a specific gauge theory, with given matter content and interactions, propagating on a certain four dimensional manifold (as defined above it is $R \times S^3$). Aren't all of those choices "backgrounds", and isn't the theory then manifestly background dependent?

Returning to  the discussion in the introduction, particularly to the distinction made between dynamical backgrounds and superselection sectors, one is required to decide which aspects of the theory are chosen by the dynamics, and which cannot be changed by any finite dynamical process. From the gravitational description, it seems clear that asymptotic boundary conditions are precisely those aspects of the theory which define "superselection sector" \footnote{For a discussion of evidence that asymptotic boundary conditions cannot be changed by dynamics, see for example \cite{Banks:2000zy}.}. More technically those boundary conditions are associated with non-normalizable modes in the gravitational descriptions of the system, and those do not fluctuate.

 This is even more clear in the gauge theory description of the system: the matter content, the Lagrangian, the rank of the gauge group and the manifold on which the (non-gravitational) theory propagates, those are all fixed for all states of the theory, and for all physical processes allowed in the theory. On the other hand, the correspondence does not specify any background metric, background values for any of the fields, or any other aspect of the theory that can change dynamically. When specifying the boundary conditions, the gauge theory description already sums the contribution of all bulk geometries (and other field configurations) which satisfy those boundary conditions, none of those backgrounds makes an appearance in the gauge theory description. In that sense the gauge theory description is as BI as any other theory in physics, including Einstein's general theory of relativity.

This correspondence is also an important example for the role of diffeomorphism invariance in quantum gravity. First, a subtlety: with specific boundary condition, there are two types of diffeomorphism: those which change the boundary conditions, and those which do not. The former (when they exist) are global symmetries, which have physical consequences and therefore must be visible in any variables chosen. The latter type of diffeomorphism (sometimes called bulk diffeomorphisms), those which fix the boundary conditions, are a redundancy of the description, they have no physical consequence and are implemented very differently depending on the variables chosen. For example, the definition of quantum gravity (in the superselection sector described by the given boundary conditions) through the dual gauge theory is diffeomorphism invariant with respect to the bulk diffeomorphisms. This is achieved not by the elaborate process of imposing constraints on some auxiliary Hilbert space, instead all variables making appearance in the gauge theory description are already diffeomorphism invariant. This is precisely what happens in the case of gauge dualities, described in the previous section, and summarized in figure 1.

\section{Conclusions and Outlook}

I'll conclude by commenting on implications of the above for future directions.

In trying to directly quantize the gravitational field, one of the main technical difficulties is imposing the constraints of diffeomorphism. This results in an intense study of those constraints, their algebra and representations, and the various ways those constraints can be implemented. On the other hand, in the quantum gravity theories defined via holography the algebra of diffeomorphisms \footnote {As explained above, I am only considering the diffeomorphisms which leave the boundary conditions unchanged, which then can be regarded as gauge redundancies.} is not a very useful tool, all the fields used in the holographic dual are singlets of diffeomorphism, the structure of the diffeomorphism algebra gives no information.

 The existence of holographic dualities redefines what one means by quantum gravitational theory. It seems that almost any theory can be regarded as quantum gravity in the sense of being diffeomorphism invariant.  A more useful definition of quantum gravity is that of a quantum system which possesses a classical limit containing Einstein's gravity (in some large semi-classical space). In that sense the four dimensional $N=4$ supersymmetric gauge theory is apparently a (five dimensional) quantum gravity theory, since in a suitable  limit it implies universal gravitational attraction between test masses.

We have seen that various instances of gauge freedom, including diffeomorphism invariance, are less fundamental than once thought. One may ask why such redundancies seem to arise generically  whenever one takes the classical limit.  As mentioned above, the reason seems to be locality, one needs to introduce gauge potentials and the resulting redundancies to make the formulation manifestly local.  Indeed, one of the mysteries of the gauge-gravity dualities, or any other holographic definition of a quantum gravitational theory, is that of bulk locality. The gauge theory contains everything one expects from a quantum gravitational theory, for example black holes forming and evaporating, in-falling observers etc. etc., albeit all this information is scrambled in a way that hides its local nature. This is intimately related to the fact that bulk diffeomorphisms are realized trivially in that language.

 Thus, the study of the mathematical structure of diffeomorphism invariance seems to do less to do with the fundamental structure of quantum gravity, and more to do with the limit in which the theory becomes semi-classical and local in the gravitational variables. Perhaps the intense study of the structure of diffeomorphism symmetry and possible semi-classical quantizations of the gravitational fields can aid in identifying the local bulk information in quantum gravity theories defined holographically. For example, it would be nice to see an attempt to provide a loop quantization of asymptotically AdS spaces. On general principles one would obtain a conformal field theory, and the relation of that CFT to the $N=4$ supersymmetric gauge theory may be very illuminating.

\section*{Acknowledgements}

I thank the organizers of the "Loops 2007" conference in Morelia, Mexico for inviting me to share my point of view with them, and for interesting questions and discussions. This work is supported by a discovery grant from NSERC of Canada.

\newpage

\bibliography{background}

 \end{document}